\documentclass[conference]{IEEEtran}
\usepackage{amsmath,amssymb,amsfonts, amsthm}
\usepackage{mathtools}
\usepackage{dsfont}
\usepackage{algorithmic}
\usepackage{algorithm2e}
\usepackage{float} 
\usepackage{afterpage}
\usepackage{graphicx}
\usepackage{textcomp}
\usepackage{xcolor}
\usepackage{tikz}
\usetikzlibrary{intersections, shapes}
\usetikzlibrary[datavisualization]
\usepackage{comment}
\usepackage{gensymb}
\usetikzlibrary{calc}
\usepackage{import}
\usetikzlibrary{decorations.pathreplacing} 
\usepackage{pgfplots}
\usepackage{pgfplotstable}
\usepgfplotslibrary{fillbetween}
\usepackage[normalem]{ulem}
\graphicspath{{images/}}
\usepackage{mdefs}
\usepackage{cite}
\usepackage{lipsum}
\usepackage{subcaption}
\usepackage[utf8]{inputenc}
\usepackage{hyperref}
\DeclareUnicodeCharacter{2061}{}
\usepackage[top=0.72in, bottom=1.0in, left=0.63in, right=0.66in]{geometry} 

\usepackage{algorithmic}
\usepackage{algorithm2e}

\floatstyle{ruled}
\newfloat{algorithm}{tbp}{loa}
\providecommand{\algorithmname}{Algorithm}
\floatname{algorithm}{\protect\algorithmname}

\def\BibTeX{{\rm B\kern-.05em{\sc i\kern-.025em b}\kern-.08em T\kern-.1667em\lower.7ex\hbox{E}\kern-.125emX}}

\def\GHz{~\mathrm{GHz}}
\def\MHz{~\mathrm{MHz}}
\def\Hz{~\mathrm{Hz}}

\def\m{\mathrm{m}}
\def\km{\mathrm{km}}

\def\s{~\mathrm{s}}
\def\dB{~\mathrm{dB}}
\def\dBm{~\mathrm{dBm}}

\def\minof#1#2{\mathrm{min}\left(#1, #2\right)}
\def\minof#1#2{\mathrm{min}\left(#1, #2\right)}

\title{Sensing of Side Lobes Interference for Blockage Prediction in Dense mmWave Networks}

\author{\IEEEauthorblockN{Mohamed Sana, Hiba Dakdouk, Benoit Denis}
\IEEEauthorblockA{CEA-Leti, Université Grenoble Alpes, F-38000 Grenoble, France\\
Email: \{mohamed.sana, hiba.dakdouk, benoit.denis\}@cea.fr}}

\newcommand{\titleheader}{This work has been accepted for publication in 2023 IEEE International Symposium on Personal, Indoor and Mobile Radio Communications (PIMRC): Track 2: Networking and MAC}

\def\ps@IEEEtitlepagestyle{%
\def\@oddhead{\mbox{}\scriptsize \titleheader \rightmark \hfil }%
}
\makeatother

\begin{document}

\maketitle

\begin{abstract}
The integration of sensing capability in the design of wireless communication systems is foreseen as a key enabler for efficient radio resource management in next-generation networks. This paper focuses on millimeter-wave communications, which are subject to severe attenuation due to blockages, ultimately detrimental to system performance. In this context, the sensing functionality can allow measuring or even imaging the wireless environment allowing anticipation of possible link failures, thus enabling proactive resource reallocation such as handover. This work proposes a novel mechanism for opportunistic environment sensing, which leverages existing network infrastructure with low complexity. More specifically, our approach exploits the fluctuations of interference, perceived in antenna side lobes, to detect local activity due to a moving blocker around the reference communication link. Numerical evaluations show that the proposed method is promising as it allows effective assessment of the blocker direction, trajectory and possibly, its location, speed, and size.
\end{abstract}


\section{Introduction}

Millimeter-Wave (mmWave) frequencies (ranging, \egn, between $28$ and $300$ GHz) are getting great attention recently for their various advantages over traditional radio frequencies (sub-6 GHz).
With the large spectrum available at these frequencies, mmWave technology can effectively boost the network capacity. 
It also supports advanced beamforming techniques, which allow for highly directional signal transmissions, reducing interference and enhancing network performance.
However, these advantages also come along with a critical challenge: mmWave communications suffer from severe path-losses and are very sensitive to blockages and attenuation by physical obstacles (\egn, buildings, trees, human body). 
Penetration losses through the human body can range between $20-40\dB$, whereas attenuation through buildings can be as high as $40-80\dB$ \cite{andrews2016modeling}.

Frequent interruptions and long-duration blockages may cause severe degradation in the quality of service (QoS) of end-users, requiring frequent handover procedures that are detrimental to network performance \cite{SanaHO2022}. 
Therefore, effective blockage prediction mechanisms are needed to enable efficient radio resource management (RRM).
Joint communication and sensing has been identified as a key feature of future 6G systems \cite{liu2022integrated}. 
These sensing capabilities could be used to improve network performance by providing optimization inputs for network steering, including the ability to detect objects that (temporarily) obstruct or block the line of sight (LoS) between two communicating nodes \cite{ericsson}. 
 Sensing the surrounding for detecting moving blockages in mmWave communications has become a fundamental research topic \cite{oguma2016performance,charan2021vision,demirhan2022radar,marasinghe2021lidar,ali2019early,alrabeiah2020deep}.
A major axis of research that aims to predict and prevent blockages in mmWave systems considers making use of in-band mmWave signal and data rate observations. 
The authors in \cite{wu2022blockage} uses the fluctuation of the received power level occurring before the shadowing event to predict the future time instance of a blockage with the aid of deep neural networks. 
However, the prediction accuracy decreases as the blocker is far from the mmWave link, which means that it can only be detected accurately when it is close to the mmWave communication beam. 
In \cite{koda2019handover}, the data rate fluctuation occurring before the shadowing indicates the potential blockage. 
Using deep reinforcement learning techniques, the authors could predict handover timings while obstacle-caused data rate degradation are predicted before the degradation occurs.
On the other hand, authors in \cite{hersyandika2022guard} propose the use of an additional passive mmWave beam (guard beam) next to the main communication beam that is intended to sense the environment by expanding the field of view of a base station (BS).
Thus, a blocker could be detected early by observing the received signal fluctuation resulting from the non line of sight (NLoS) component from the user equipment (UE) due to the  blocker's presence within the field of view. 
Yet, all these approaches are limited in terms of detection range, as the blocker should be close to the main communication beam causing fluctuation in the received signal, and might fail with large-velocity moving blockers. In addition, the sensing feature may require an additional and dedicated mechanism (\egn, a dedicated beam). 

In contrast, we propose a mechanism that exploits side lobes information for the passive and opportunistic sensing of a dense mmWave network. Network densification is a key feature of future networks that will further improve their capacity \cite{kamel2016ultra}. 
At the same time, it may also lead to increased intra- and inter-cell interference, which may impact communication performance.
However, in this work, we take advantage of this specific characteristic of dense networks to detect the presence of moving blockers in the surrounding environment of communicating nodes.
Our method relies on the observation of the interference fluctuations in antenna side lobes caused by the existence of moving blockers in angular sectors around the communication link of concern. Unlike the aforementioned studies, our approach is capable to detect and track moving objects all around the sensing device, \ie in range of $360^{\degree}$.
This makes it possible to predict some characteristics of blockers, including their trajectory and velocity, allowing early detection of blockage events and avoiding link outages by triggering, \egn, a handover procedure.

\section{System Model}
\label{sec:system_model}
We consider a dense mmWave network composed of a set $\mathcal{B}=\{b_1,...,b_M\}$ of $M$ BSs deployed in a bi-dimensional Euclidean space of radius $R$ to provide service coverage to a set $\mathcal{U}=\{u_1,...,u_K\}$ of $K$ UEs. 
We assume BSs and UEs form two distinct homogeneous Poisson Point Processes (PPP) with densities $\lambda_b~[\m^{-2}]$ and $\lambda_u~[\m^{-2}]$ respectively such that in average, $\mathbb{E}[M] = \lambda_b \pi R^2$ and $\mathbb{E}[K] = \lambda_u \pi R^2$. 
In this dense network, a mobile and passive object (\egn, a robot), modelled as a cylindrical object of radius $r_B$ moves around, causing the blockage of interfering and direct communication paths. 
Let $\omega_B(t)$ denote its angular velocity and $d_B(t)$ its distance with respect to (\wrtn) BS $b_0$, referred to as the \emph{typical BS} and taken as the reference point in the following. Clearly, $(\omega_B(t), d_B(t))$ characterizes the trajectory of the blocking object. 
In this work, we propose a novel approach for a passive sensing of such a moving object, partially identifying its trajectory by leveraging the interference perceived in the side lobes of the antenna radiation patterns.

We focus on an uplink setting with only LoS communications for both direct and interfering links. In this scenario, an initial access phase allows new UEs to perform beam training and alignment mechanisms, configuring the appropriate beams, which exploit the maximum directivity gain \wrt serving BSs for the service phase. 
For simplicity, we assume each UE gets associated with the closest BS, as we do not specifically address the user association problem. 
However, this problem can be efficiently solved using approaches proposed in \cite{Sana21} to optimize service coverage. During the service phase, BSs exploit the fluctuations of interference perceived in their antennas side lobes, resulting from simultaneous communications with UEs, for sensing their nearby environment to detect blockages.  

\begin{figure}[!t]
    \centering
    \includegraphics[width=\columnwidth]{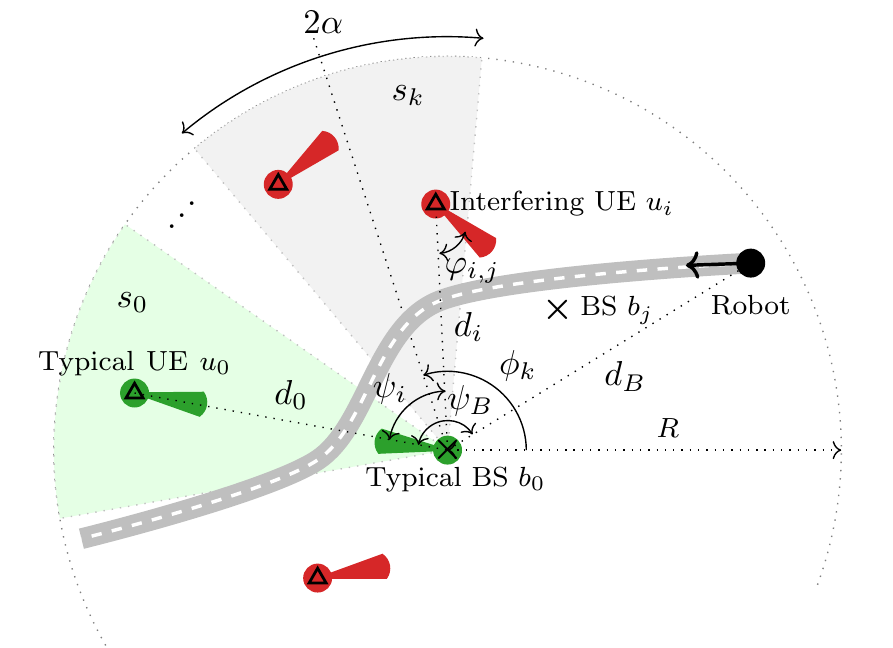}
    \caption{Network model with 3 UEs interfering on a communication between a typical UE $u_0$ and serving BS $b_0$. In this network, a mobile robot moves around causing blockages.}
    \label{fig:interference-model}
\end{figure}

\vspace{0.3em}
\noindent
\textbf{Antennas.} In our system model, UEs and BSs are equipped with antenna arrays to perform directional beamforming. 
For easy analysis, we assume that antenna arrays operate on the same elevation plane, and accordingly, we set the beam elevation angle to zero. 
Therefore, we denote with $G^{\rm Tx}_{\theta}(x)$ and $G^{\rm Rx}_{\vartheta}(x)$ the transmitter and the receiver 2D antenna radiation pattern respectively, where $\theta$ and $\vartheta$ are the beam width and, $x$ is the azimuth angle to the main lobe (either $\psi$ or $\phi$ in \fig{fig:interference-model}). For the tractability of analysis, we approximate the actual 2D array patterns with a sectored Gaussian directional antenna model \cite{Yang2018} whose beamforming gain is given as follows:
\begin{align}
\label{eq:antenna_model}
    G_z^{\ell}(x) = \left \{
				\begin{array}{l l}
					G_{m}^{\ell} e^{-\rho_z x^2}, &\quad \text{if } |x| \leq \frac{z}{2},\\
					G_{s}^{\ell}, &\quad \text{otherwise},
				\end{array},~z\in\{\theta, \vartheta\},
			\right.
\end{align}
where $\rho_z=2.028 \dfrac{\ln⁡{(10)}}{z^2}$ and $z$ is the beam width. 
In addition, $G_m^{\ell}$ and $G_s^{\ell}$ denote the gain of the main lobe and the side lobes as per $\ell\in\{\mathrm{Tx}, \mathrm{Rx}\}$, respectively. 
Following these definitions, we define the antenna peak-side-lobe (PSL) gain as ${\mathrm{PSL}^{\ell} = {G_m^{\ell}}({G_s^{\ell}})^{-1}}$.
In particular, the value of $\mathrm{PSL}^{\ell}$ depends on the number of antenna elements.

\noindent
\textbf{Propagation Channel.} We adopt the commonly used Friis propagation loss model \cite{bai2014coverage}, where the received power $P^{\mathrm{Rx}}$ is given as a function of the transmitted power $P^\mathrm{Tx}$ and the distance $d$ between the transmitter and the receiver:
\begin{equation}\label{eq:channel-model}
    P^{\mathrm{Rx}}(t) = \chi(t) \zeta(t) P^{\mathrm{Tx}} G_{\theta}^{\mathrm{Tx}}(\varphi) G^{\rm H}(d) G_{\vartheta}^{\mathrm{Rx}}(\psi),
\end{equation}
where $\varphi$ and $\psi$ represent the azimuth angles at the transmitter and receiver respectively. 
Here, $G^{\rm H}(d)$ is the distance-dependent channel gain, which captures the effect of path-loss and large-scale shadowing as follows:
\begin{equation}\label{eq:channel-gain}
    G^{\rm H}(d)|_{\text{dB}} = \mathrm{PL}_0 + 10 \eta \log_{10}⁡\left(\frac{d}{d_{\rm ref}}\right)+X_{(\sigma_s)}, 
\end{equation}
where $\mathrm{PL}_0$ denotes the pathloss constant, $d_{\rm ref}$ is a reference distance, $\eta$ denotes the pathloss exponent and $X_{(\sigma_s )}$ represents the static shadowing effect, modeled as a Gaussian variable with zero mean and variance $\sigma_s^2$. 
Also, $\zeta(t)$ represents the small-scale fading coefficient, which follows a $m$-Nakagami distribution. Eventually, $\chi(t)$ denotes the shadowing effect due to the passive object moving around the corresponding link.
We adopt the following Gaussian modeling as in \cite{qi2017measurements}:
\begin{align}
    \chi(t)|_{\rm dB} = -A\exp{\left(-\frac{|\psi - \psi_B(t)|^{2}}{\sigma_B^2}\right)} 
\end{align}
where $\psi_B(t)$ is the relative angle between the blocker and the receiver main lobe. Accordingly, $|\psi - \psi_B(t)|$ represents the angle between the blocker and the interfering link with relative angle $\psi$ \wrt the receiver (see \fig{fig:interference-model}); $A$ denotes the attenuation (in dB) when the link is fully blocked (\ien,~$|\psi - \psi_B(t)|=0$) and {$\sigma_B$ is a parameter that depends on the characteristics of the blocker (\egn,~size) and radio parameters}. Although simplistic, this model allows an effective geometric analysis of blocking phenomena.

\vspace{0.3em}
\noindent
\textbf{Cell interference.}
In our system model, as we do not specifically optimize beamforming, interference results from the overlapping of communication beams between mmWave BSs and UEs. 
Indeed, let us consider a typical BS $b_0$ placed at a distance $d_0$ from its served UE $u_0$. 
We refer to $u_0 \rightarrow b_0$ as the reference link. 
An interfering UE $u_i$, located at a distance $d_i$ with a relative angle of arrival (AoA) $\psi_i$ \wrt $b_0$, is served by another BS $b_j$ in a relative angle of departure (AoD) $\varphi_{i,j}$ (see \fig{fig:interference-model}). 
We denote with $I_{i,j}$, the resulting interference perceived by BS $b_0$: 
\begin{align}
    I_{i,j}(t) &= \chi_{i}(t) \zeta_i(t) P_i^{\mathrm{Tx}} G^{\rm Tx}_{\theta}(\varphi_{i,j}) G^{\rm H}(d_i) G^{\rm Rx}_{\vartheta}(\psi_i)
\end{align}
Thus, the total interference perceived by the typical BS as a function of signal angle of arrival (AoA) $\psi$ reads as:
\begin{align}
    I(\psi, t)=  \sum_{\mathclap{\substack{(u_i,b_j)\in\mathcal{U}\times\mathcal{B}\backslash(u_0, b_0)}}}  I_{i,j}(t) \delta(\psi-\psi_i), \label{eq:interf-inter}
\end{align}
where, $\delta(\cdot)$ is the Dirac delta function.

\section{Sensing of side lobes interference}
\label{sec:SLS}
\subsection{On the design of the sensing matrix}
As we assume spatial reuse of the spectrum across the network, interference may come from multiple directions, especially in dense networks.
We assume that BSs can sense the sum-interference perceived in their side lobes from a certain sector. Typically, we define:
\begin{align}
    I_{s(\phi, \alpha)}(t) = \int_{s(\phi, \alpha)} I(\psi, t)\mathrm{d}\psi,
\end{align}
where $s(\phi, \alpha)$ denotes sector of angular wide $2\alpha$ with absolute orientation $\phi$ \wrt to horizontal axis as shown in \fig{fig:interference-model}.

We introduce a novel metric, the signal-to-sector-interference-plus-noise ratio,\footnote{In practice, $\gamma_{s(\phi, \alpha)}$ can be estimated using approaches similar to beam- or angular-domain channel estimation techniques \cite{Gao2015, Zhao2020}. Except here, there is no need to estimate channel state information but rather to capture integrated interference power in each beam. Obviously, the time spent to acquire the full sensing matrix must be compliant with blocker mobility dynamics to guarantee observability. These temporal aspects of the sensing procedure are out of the scope of this paper, but left for future works.} denoted $\gamma_{s(\phi, \alpha)}$, between the typical UE $u_0$ and its serving BS $b_0$, which we define as:
\begin{align}\label{eq:SINR}
    \gamma_{s(\phi, \alpha)}(t) = \frac{P_0^{\mathrm{Tx}}\chi_0(t) \zeta_0(t) G_{m}^{\mathrm{Tx}} G^{\rm H}(d_0) G_{m}^{\mathrm{Rx}}}{I_{s(\phi, \alpha)}(t)  + N_0B}.
\end{align}
Here $B$ is the bandwidth, and $N_0$ is the noise power spectral density. 

Now, let divide the $2\pi$ angular space into $n+1$ non-overlapped sectors of angle $2\alpha=\frac{2\pi}{n+1}$, here preferably chosen contiguous to cover all 2D space. We refer to $s_k$ as the $k$-th sector with absolute orientation $\phi_k$, where $k=0$, correspond to the sector whose orientation coincides with the link $(u_0 \rightarrow b_0)$. Thus, $\phi_k = \phi_0-2k\alpha$. Finally, let us consider a short-term observation window of size $\tau+1$ and let define: 
\begin{align}
\label{eq:sensingMatrix}
\boldsymbol{\Lambda}_{\tau,n}(t) = 
\begin{pmatrix}
\gamma_{s_0}(t) & \dots & \gamma_{s_n}(t)\\
\gamma_{s_0}(t-1) & \dots & \gamma_{s_n}(t-1)\\
\vdots  & \ddots & \vdots\\
\gamma_{s_0}(t-\tau) & \dots & \gamma_{s_n}(t-\tau)
\end{pmatrix}
\end{align}
We refer to $\boldsymbol{\Lambda}_{\tau,n}(t)$ as the sensing matrix. 
The idea behind defining such matrix is that, the blockage of an interfering link around the typical BS in the sectorized regions induces fluctuations of the values of $\gamma$ whose intensity depends on the shadowing $\chi$, and on the side lobe gains (more specifically on the PSL of the antenna). By sensing the side lobes interference, we show that early detection of blockage events is possible, thus allowing efficient resource reallocation. 

Clearly, the accuracy of the detection also depends on sector angular width $\alpha$ and the density of interferers (here UEs). Indeed, the probability to have at least one interferer in sector $s_k(\alpha)$ is given by:
\begin{align}
\label{eq:interferer_proba}
    p_1 = 1 - e^{-\lambda_u \alpha R^2}.
\end{align}
Thus, the smaller the value of $\alpha$, the higher the resolution of matrix $\boldsymbol{\Lambda}$. At the same time, the smaller is $\alpha$, the scarcer the matrix $\boldsymbol{\Lambda}$ as $p_1 \rightarrow 0$. 

\subsection{Singular Value decomposition of the sensing matrix for blocker signature detection}
The sensing matrix can be viewed as the sum of three components:
\begin{align}
    \boldsymbol{\Lambda}_{\tau,n}(t) = \boldsymbol{\Lambda}_{\tau,n}^{(\circ)}(t) + \boldsymbol{\Lambda}_{\tau,n}^{(\bullet)}(t) + \boldsymbol{\Lambda}_{\tau,n}^{(\rm noise)}(t),
\end{align}
where, $\boldsymbol{\Lambda}_{\tau,n}^{(\circ)}(t)$ denotes the blockage-free sensing matrix, $\boldsymbol{\Lambda}_{\tau,n}^{(\bullet)}(t)$ denotes the blockage signature matrix, and $\boldsymbol{\Lambda}_{\tau,n}^{(\rm noise)}(t)$ is the additive noise due to \eg, random fading, scattering, etc. 
Our objective is thus to capture the moving blockage signature, \ien, $\boldsymbol{\Lambda}_{\tau,n}^{(\bullet)}(t)$. To do so, we apply a blind source separation technique based on singular value decomposition (SVD), similar to idea used for processing seismic data \cite{vrabie2004modified}. Following this idea, we decompose the sensing matrix $\boldsymbol{\Lambda}_{\tau,n}(t)$ into the product of three matrices as follows:
\begin{align}
    \boldsymbol{\Lambda}_{\tau,n}(t) = \mathrmbold{U}\mathrmbold{\Sigma}\mathrmbold{V}^T
\end{align}
where $\mathrmbold{U}\in\mathbb{R}^{n\times r}$, $\mathrmbold{V}\in\mathbb{R}^{r\times \tau}$, and $\mathrmbold{\Sigma}\in\mathbb{R}^{r\times r} = \mathrm{diag}(\sigma_1, \sigma_2, \dots, \sigma_r)$, where  $r = \mathrm{rank}(\boldsymbol{\Lambda}) \leq \minof{\tau}{n}$ (the number of non-zero singular values) and $\sigma_1 \geq \sigma_2\geq \dots\geq \sigma_r > 0$ are the singular values. This decomposition can be further written in the following form:
\begin{align}\label{eq:svd}
    \boldsymbol{\Lambda}_{\tau,n}(t) &= \sum_{l=1}^r \sigma_l \mathrmbold{u}_l \mathrmbold{v}_l^T \\ \nonumber
    &=  \underbrace{\sum_{l=1}^{l_0} \sigma_l \mathrmbold{u}_l \mathrmbold{v}_l^T}_{\boldsymbol{\Lambda}_{\tau,n}^{(\circ)}(t)} + \underbrace{\sum_{l=l_0+1}^{l_1} \sigma_l \mathrmbold{u}_l \mathrmbold{v}_l^T}_{\boldsymbol{\Lambda}_{\tau,n}^{(\bullet)}(t)} + \underbrace{\sum_{l=l_1+1}^{r} \sigma_l \mathrmbold{u}_l \mathrmbold{v}_l^T}_{\boldsymbol{\Lambda}_{\tau,n}^{(\rm noise)}(t)},
\end{align}
where $\mathrmbold{u}_l$ (resp. $\mathrmbold{v}_l$) are the columns of the semi-unitary matrix $\mathrmbold{U}$ (resp. $\mathrmbold{V}$) and $1\leq l_0\leq l_1\leq r$.
As for seismic data processing, the mobility of the blocker waves the singular values. 
Thus, a proper selection of $l_0$ and $l_1$ allows a full capture of the blockage signature matrix. In practice, the value of $l_0$ and $l_1$ can be chosen by finding abrupt changes in the slope of the curve of relative singular values.

\section{Numerical results}
\label{sec: analysis}

We consider a network of $M$ BSs and $K$ UEs distributed in the space of a circular industrial environment of radius $R = 100~ \m$ according to homogeneous PPP with densities  $\lambda_b = 6\times10^{-4} ~\m^{-2}$ and $\lambda_u=1.5\times10^{-3} ~\m^{-2}$ respectively.
Each UE is associated to its closest BS.
A BS $b_0$ located at the center of the network is taken as a reference: it performs side lobe sensing and computes the detection matrix on a link with a randomly chosen associated UE $u_0$.
We set the width of the angular sectors to $2\alpha=10\degree$ (\ie $n+1=36$ sectors overall, covering the 2D space), and the size of the observation window is set to $\tau=50\s$. 
The sector width is always set equal to the antenna beamwidth $\vartheta$ of BS $b_0$.
When we detect activity in a sector, the estimated angular position $\hat{\psi}_B$ of the moving object is taken equal to the orientation of the sector resulting in a mean approximation error of $5\degree$ (if the object is actually there). 
We consider the pathloss model in Eq. \eqref{eq:channel-gain}, where the values of its parameters have been characterized via in-lab experiments based on mmWave channel sounding in a representative industrial Internet of Things (IoT)-type scenario \cite{mudonhi2022mm}.
Other simulation parameters are presented in Table \ref{tab:sim_params}.

\subsection{Detection of blocker signature} 
\begin{figure*}[]
  \centering
  \begin{subfigure}[t]{0.28\textwidth}
    \includegraphics[width=\textwidth]{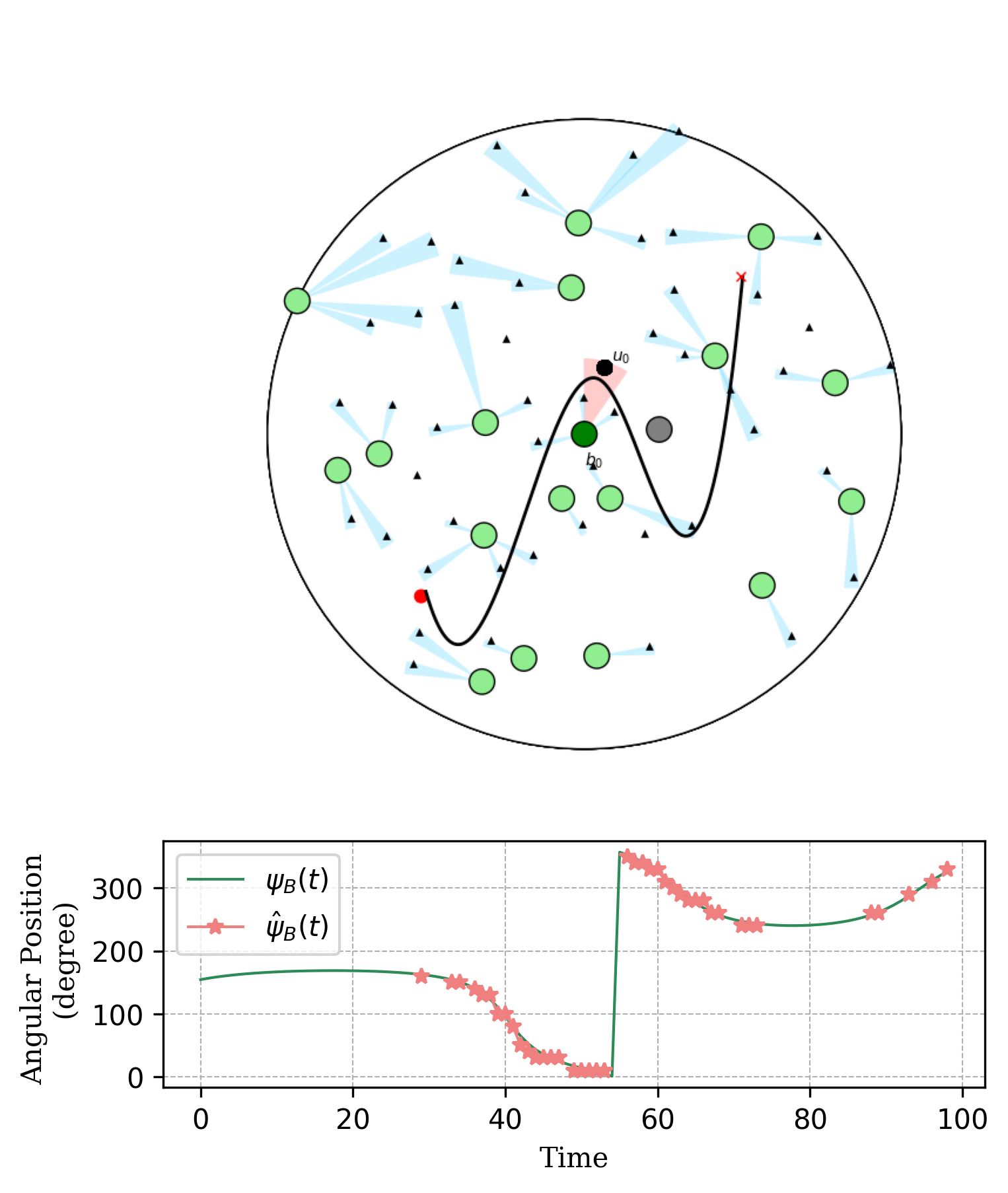}
    \caption{Deployment and blocker trajectory.}
    \label{fig:deployment}
  \end{subfigure}
  \begin{subfigure}[t]{0.33\textwidth}
  \centering
    \includegraphics[width=\textwidth]{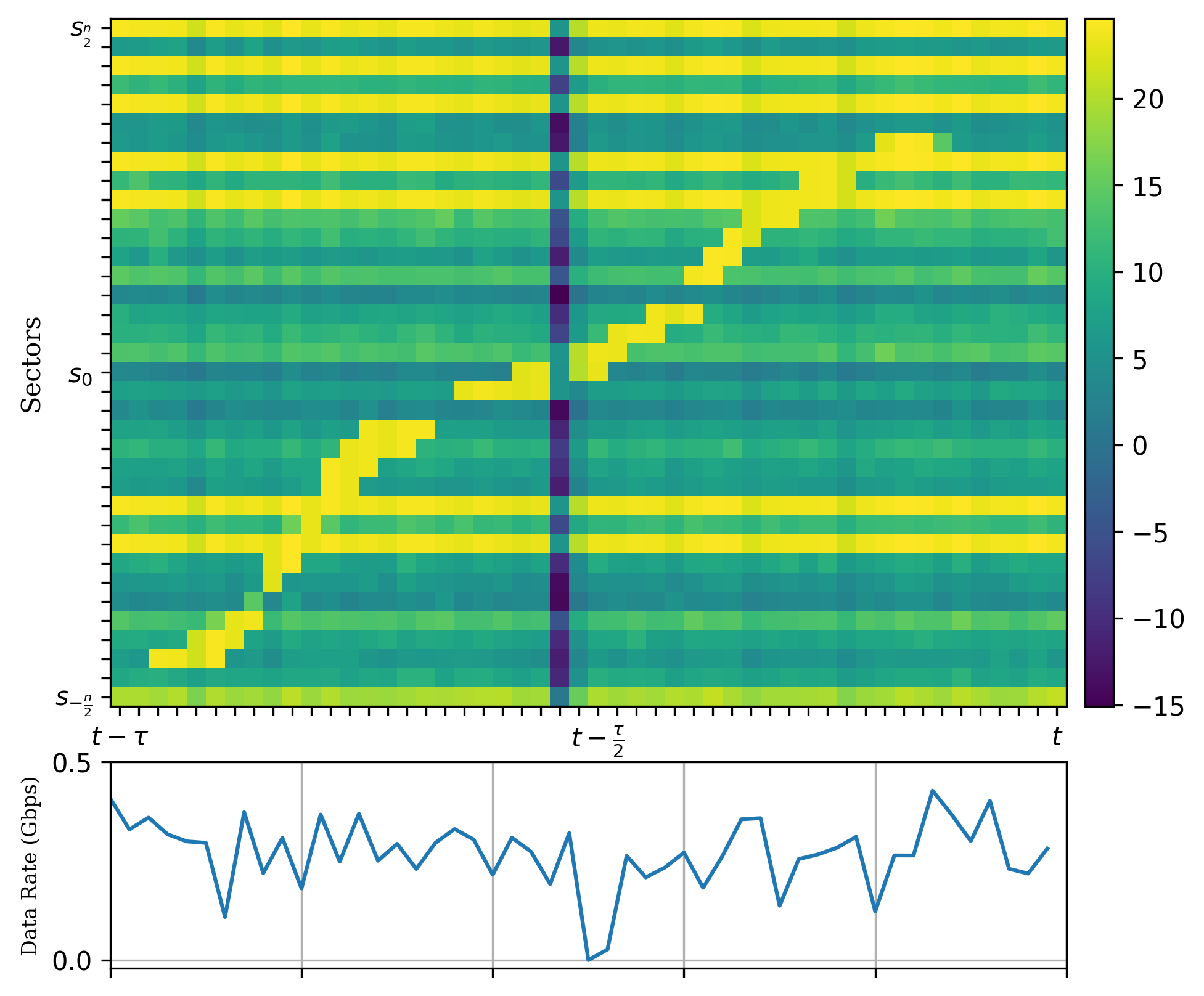}
    \caption{Raw sensing matrix $\boldsymbol{\Lambda}_{\tau,n}$ (in log scale).}
    \label{fig:rawMatrix}
  \end{subfigure}
  \begin{subfigure}[t]{0.33\textwidth}
    \centering
    \includegraphics[width=0.95\textwidth]{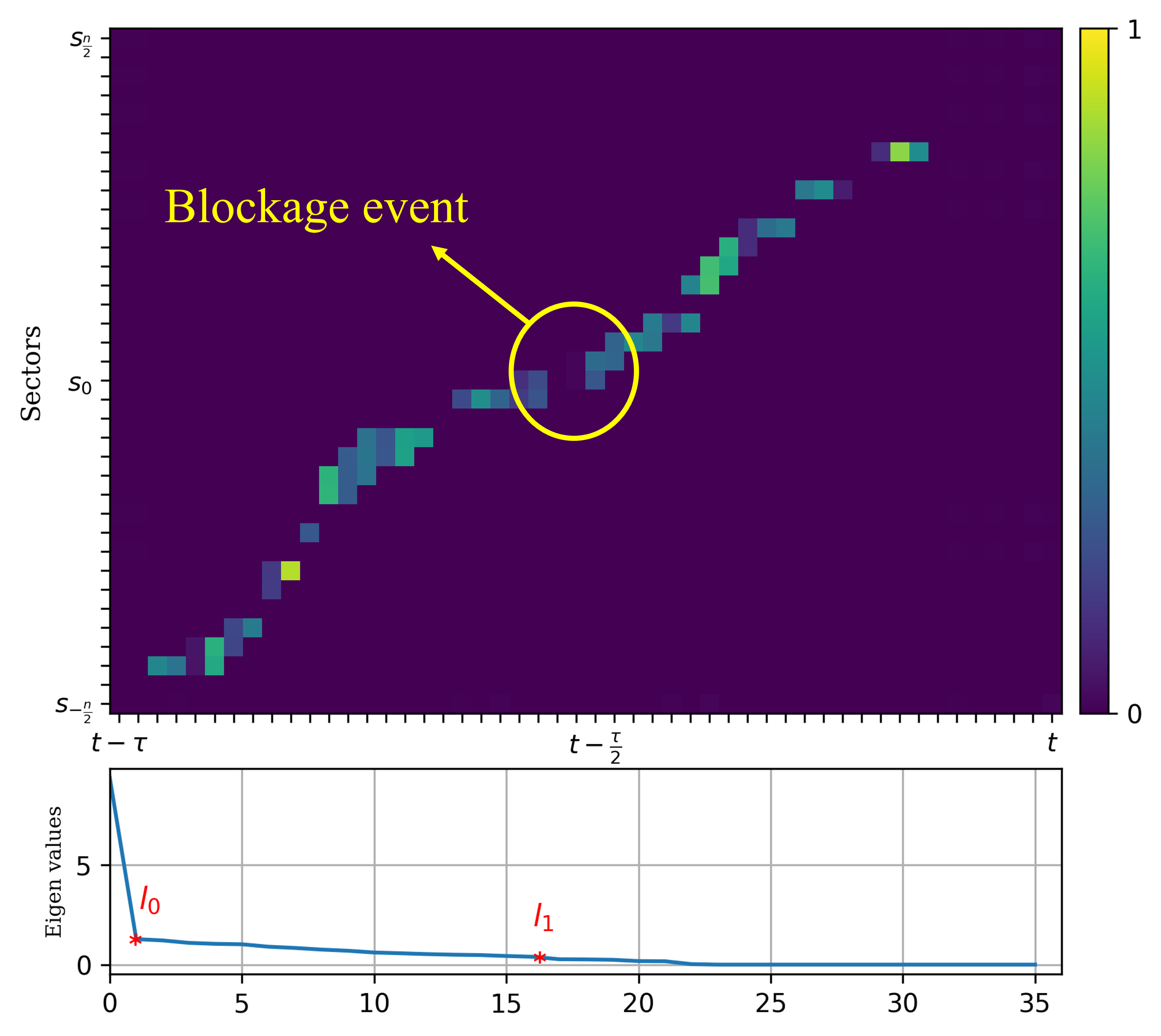}
    \caption{Blocker signature.}
    \label{fig:signature}
  \end{subfigure}
  \caption{Example of blockage detection using side lobe sensing mechanism.}
  \label{fig:detection_example}
\end{figure*}

For a random network deployment, we consider a mobile object of radius $r_B=1~\m$ moving along a random trajectory with an arbitrary velocity ($w_B \sim [-15\degree, 15\degree]s^{-1}$). Figure \ref{fig:deployment} shows an example of a network deployment, where multiple deployed BSs jointly provide service to UEs. In this figure, a blocker is moving, following a trajectory, which crosses the communication link $u_0\rightarrow b_0$ at a certain point in the main sector $s_0$. Figure \ref{fig:rawMatrix} presents the logarithm of the raw sensing matrix; $\log(\boldsymbol{\Lambda}_{\tau,n})$ of Eq. \eqref{eq:sensingMatrix}, before being processed. The processing applied to the sensing matrix makes it possible to effectively reveal the signature of the mobile object by empirically setting $l_0=1$ and $l_1=17$ in Figure \ref{fig:signature}, {where the color of each pixel quantifies the strength of the signature, \ie the likelihood of the blocker being present in the corresponding sector.} 
Thanks to this processing, it is possible to detect earlier the occurrence of the blockage, \ie as the blocker approaches sector $s_0$, and therefore to be able to cope with it before it happens. Besides, it is also possible to get information on the blocker trajectory as the bottom of Figure \ref{fig:deployment} reveals.
This figure compares the estimated trajectory of the blocker to its actual one in terms of its angular position \wrtn the link $u_0\rightarrow b_0$. 
We can observe that the proposed method allows an effective detection of the blocker and a follow-up of its angular trajectory. 
Note, however, that due to the low number of UEs in some regions of the network, which are the interferers on which this approach depends on to estimate the position of the blocker, the detection is not effective at all times (\ien, lack of observations regarding the blocker for some time epochs). 
Nevertheless, the detection is still accurate in the main region of interest, \ie when the blocker is close to the BS as shown in Figure \ref{fig:deployment}, that makes it possible to anticipate a blockage event. 
Beyond, both the missed detection events and the observation quantization error caused by space sectorization (see the step-wise fluctuations of at most $5\degree$ around the ground-truth angular trajectory in Figure~\ref{fig:deployment}) could be easily overcome by standard Bayesian filtering tools, such as Extended Kalman Filtering (EKF), even if this extra processing step does not fall in the scope of this study here.
In general, the closer the blocker gets to the BS and the closer it gets to a dense region of the network, the better the detection will be.
{Obviously, from Figure \ref{fig:signature}, it is also possible to extract additional information on the direction and speed of the blocker.}
Besides, although in this paper the trajectory is detected in terms of angular position, it is possible to be more precise, and even to locate the blocker, if we consider the sharing of sensed information between different entities in the network, which will be addressed in future work.

\begin{table}[!t]
\caption{Simulation parameters}
\centering
\begin{tabular}{c|c}
\hline
Parameters & Values \\
\hline
Carrier frequency & $28~ \GHz$\\ 
Bandwidth $\mathcal{B}$ & $400~ \MHz$\\
Pathloss & $60.1 + 14 \log(d~[\km])$\\
Power $P^{\mathrm{Tx}}$ (UE) / $P^{\mathrm{Rx}}$ (BS) & $19.6/33~ \dBm$\\
Noise power spectral density $N_0$ & $-174 \dBm/\Hz$ \\ 
Small-scale fading $\sim m$-Nakagami & $m=3$\\
{\rm Rx} beamwidth $\vartheta$ & 10°\\
{\rm Tx} beamwidth $\theta$ & 135°\\
$G_0(z)$& $\pi(21.32 z + \pi)^{-1}$ \protect\cite{Yang2018} \\ 
 $G_s^{\rm Rx}$& $G_0(\vartheta)$\\
 $G_m^{\rm Rx}$ &$G_0(\vartheta)\times10^{2.028}$  \\
$G_s^{\rm Tx}$& $0$\\
 $G_m^{\rm Tx}$ &$2G_0(\theta)\times10^{2.028}$  \\
Blockage Attenuation $A$ & $100\dB$\\
$\sigma_B$ & $\sqrt{8}~ r_B$
\end{tabular}
\label{tab:sim_params}
\end{table}


\subsection{Detection accuracy vs antenna PSL}
In this section, we assess the performance of the proposed approach for blocker signature detection through $N=500$ Monte-Carlo simulations. To avoid cumbersome computations, we assume that the circular area around $b_0$ is partitioned into a mesh grid $\mathcal{G}$ consisting of equal-size cells of angle equals to $10\degree$ and radius equals to $5~\m$. We consider a passive mobile object of radius $r_B=1~\m$ moving around $b_0$ in a circular motion at different distances up to a maximum distance of $50~\m$.
The blocker moves from one cell to another in a sequential manner. For a quantitative evaluation of the detection accuracy, we adopt the following weighted mean absolute error (wMAE):
\begin{align}
    \mathrm{wMAE} = \frac{1}{N}\sum_{k=1}^N \mathbb{E}_{(d,\psi)\in \mathcal{G}}\left[w(d_B) \abs{\psi_B - \hat{\psi}_B}\right],
\end{align}
where $\hat{\psi}_B$ is the blocker estimated angular position when its actual location is in cell $(d_B, \psi_B)$ of the grid $\mathcal{G}$. Here, $w(d_B)$ is a weighted factor, which depends on the actual distance of the blocker from the sensing BS $b_0$. This allows for a realistic evaluation of the errors since the farther the blocker is, the more complex it is to predict its angular position accurately due to path losses. In practice, following the pathloss model, we define $w(d_B) = \exp(-\mu d_B^\eta)$, where $\eta$ is the path loss coefficient and $\mu$ is a scaling factor. In particular, $\mu=0$, corresponds to the non-weighted MAE. 

\begin{figure}
  \centering
  \begin{subfigure}[]{0.5\textwidth}
  \centering
    \includegraphics[width=0.95\columnwidth]{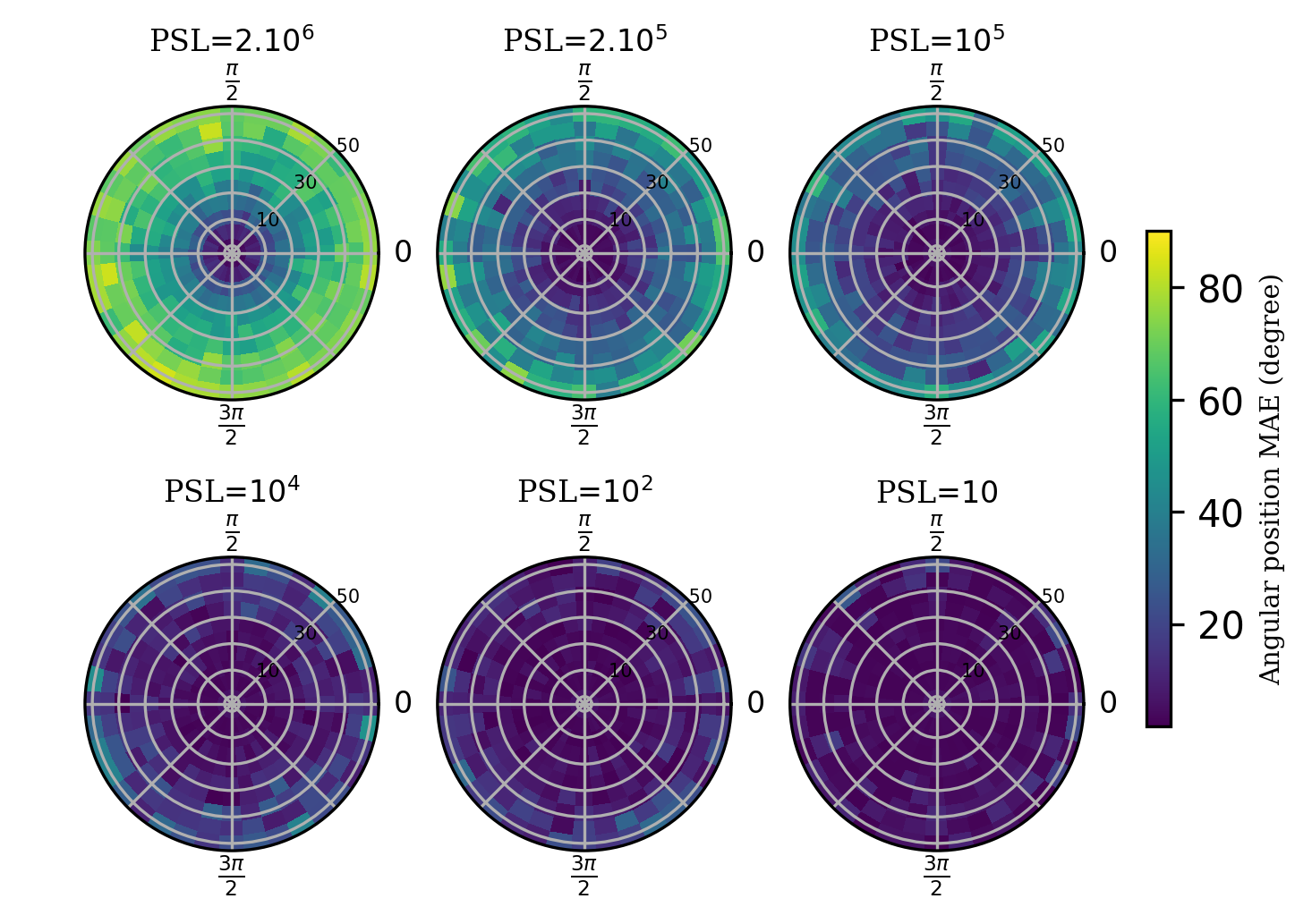}
    \caption{MAE distribution over the network area. Here $\mu=0$.}
    \label{fig:maeVSpsl_a}
  \end{subfigure}
  \vfill
  \begin{subfigure}[]{0.45\textwidth}
  \centering
    \includegraphics[width=0.9\columnwidth]{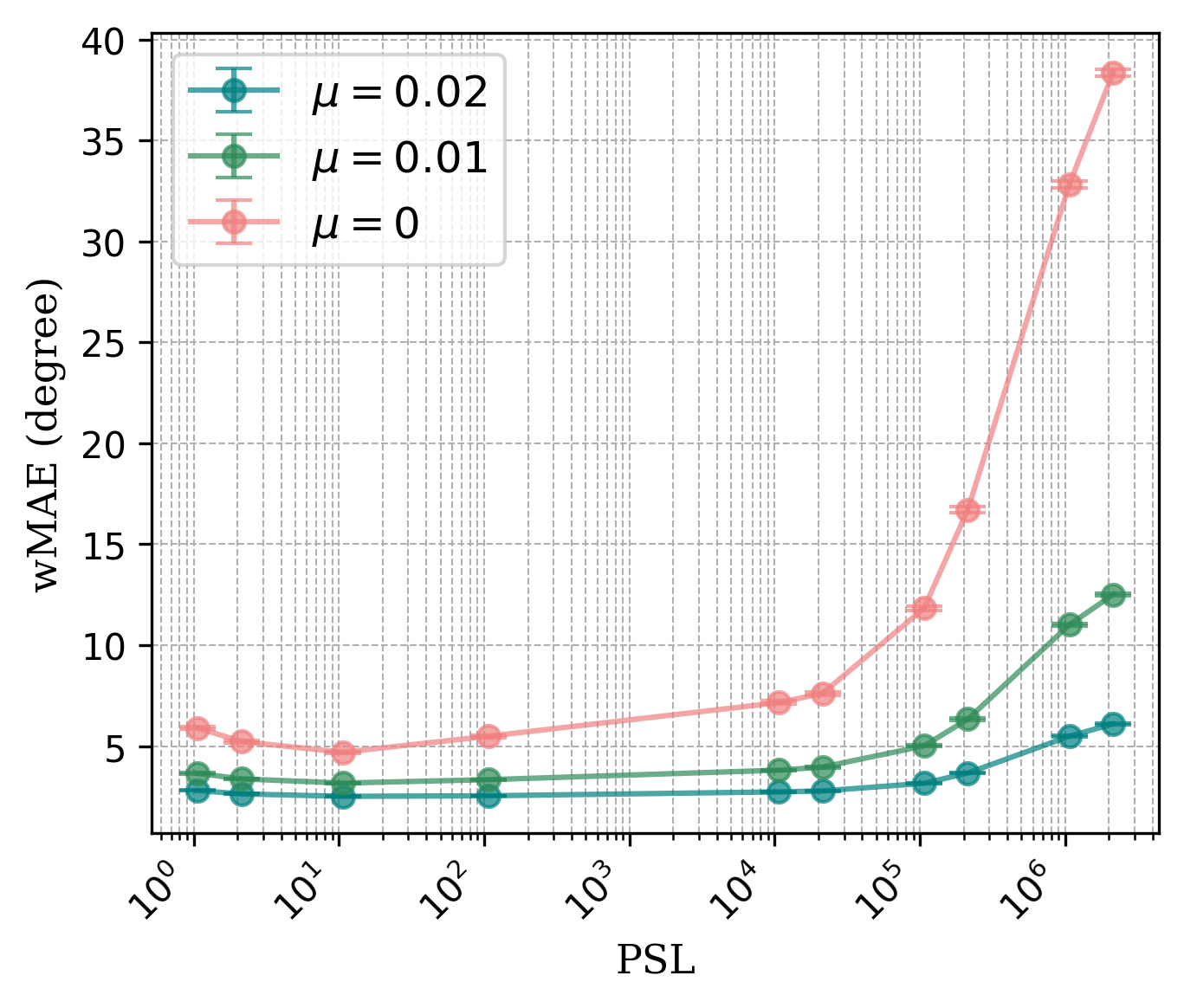}
    \caption{Angular position error, $\mathrm{wMAE}(\mu)$.}
    \label{fig:maeVSpsl_b}
  \end{subfigure}
  \caption{Weighted mean absolute error of blocker angular position vs antenna peak-side-lobe gain.}
  \label{fig:mae_vs_psl}
\end{figure}

We start by assessing the impact of antenna PSL (by varying the side lobe gain) on the MAE. 
The associated results are presented in Figure \ref{fig:maeVSpsl_a}. First, we can notice  that for any value of PSL the accuracy degrades as the blocker moves away from the center where $b_0$ is located.  
This is due to two main reasons: $1$) as the blocker moves away from the center, it is less likely to be obstructing the LoS between $b_0$ and the interfering UEs, $2$) the received power of the interference degrades as the interferer is farther, so it is highly probable that blocking far interference will not be observed by $b_0$. This result confirms the outcome of the previous experiment. In \fig{fig:maeVSpsl_b}, we show the resulting $\mathrm{wMAE}(\mu)$ for different values of $\mu$, along with the corresponding $95\%$ confidence intervals with different values of PSL. 
We can notice that the detection accuracy degrades as the PSL value increases. 
Indeed, with larger values of PSL, the interference perceived by the side lobes is weaker, and thus the detection of the signature of the blocker is less effective. 
For low PSL values, \ie when the side lobe gain approaches the main-lobe gain, the detection accuracy decreases as well because of low signal-to-interference-and-noise ratio (SINR) values in the sensing matrix. 
Thus, a trade-off can be found between the accuracy of the detection \ie by reducing the PSL, and the quality of the communication \ie by increasing the PSL to less suffer from interference.
Yet, we can observe that the proposed mechanism is still highly accurate as  $\mathrm{wMAE}(\mu=0.02) < 5\degree$  (half the sector width) and $\mathrm{wMAE}(\mu=0.01) < 10\degree$  for most PSL values.
Even the non-weighted MAE is less than $10\degree$ except for very low side lobe gain (\ie very high PSL). 


\begin{figure}[!t]
\centering
  \includegraphics[width=0.5\textwidth,height=6cm]{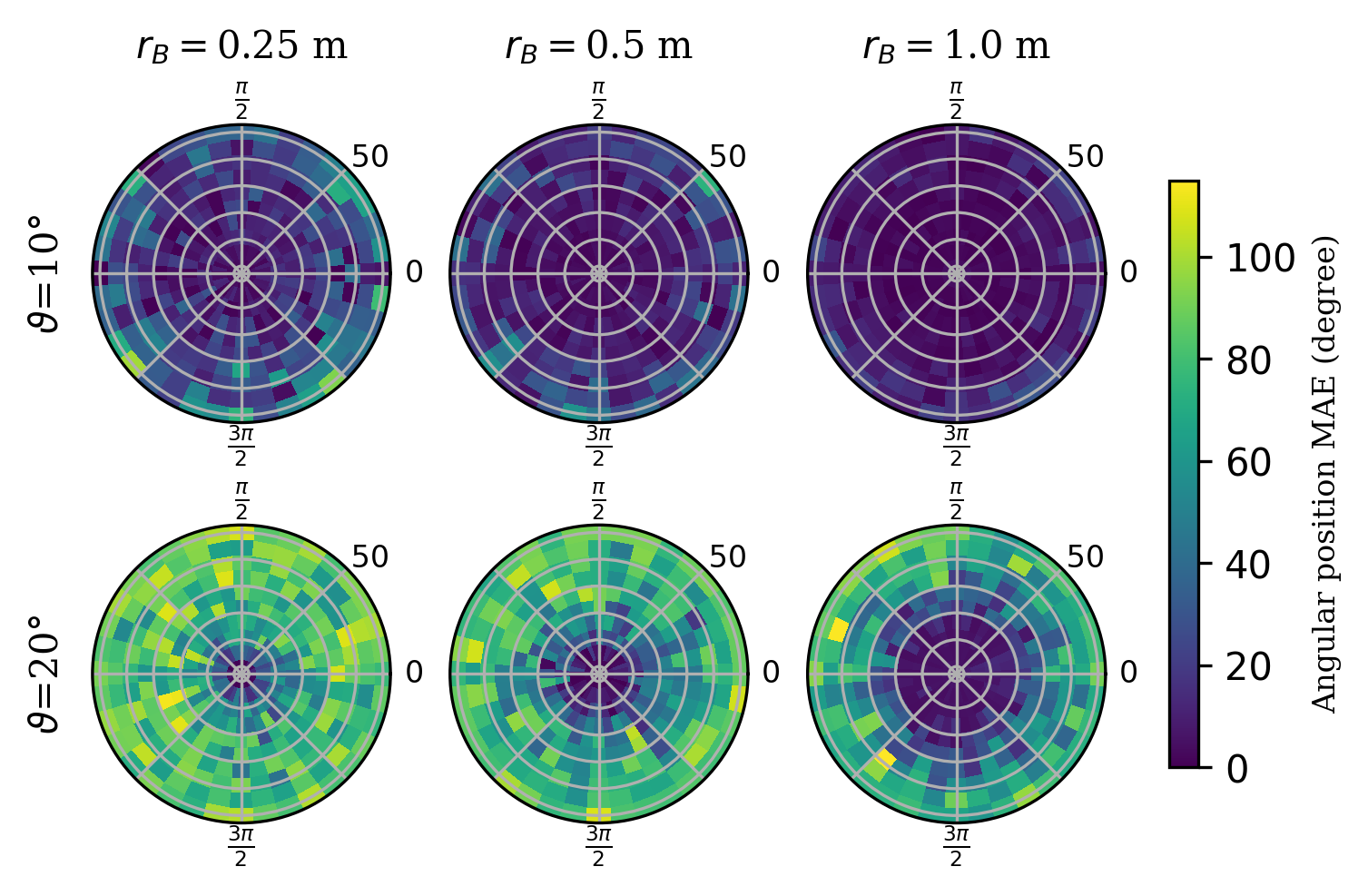}
  \caption{Mean absolute error of blocker angular position vs antenna beamwidth and blocker size. }
  \label{fig:mae_vs_bw_size}
\end{figure}


\subsection{Detection accuracy vs beamwidth and blocker size}
Other metrics that impact the performance of the side lobe sensing are the antenna beamwidth and the blocker size. 
Similar to the previous experiment, Figure \ref{fig:mae_vs_bw_size} presents the blocker angular position MAE in each cell of the network for different combinations of beamwidth and blocker radius $r_B$. 
As the blocker size increases, it gets more detectable, as it is more capable of blocking interference coming from different angles, but this indeed weakens the accuracy especially when the blocker is large and close to the center, as it could coexist in multiple sectors simultaneously. 
Also, as the beamwidth increases and so the sector width, the accuracy degrades since the range of error in a sector increases, however, the blocker gets more tracked as the probability to have interferers in the sector increases (see Eq. \eqref{eq:interferer_proba}).
Consequently, the size of the sector should be carefully tuned so that to find a trade-off between the accuracy and the continuity of detection.


\section{Conclusion}
\label{sec:conclusion}

In the context of predicting and avoiding blockages in mmWave systems, this paper presents a new mechanism that detects moving objects in the surrounding of a particular communication link. 
This approach senses the radio environment using side-lobe information in order to detect moving objects. 
It relies on observing the fluctuation in the SINR values caused by the presence of the blocker in angular sectors around the communication link of interest. 
We show that it is capable of detecting moving objects in a range of $360\degree$, without requiring any additional system unlike other reported methods.
This indeed further provides information on the position, direction, trajectory and certainly the velocity of a moving object. 
Using this information, the node would have enough time to handover to another available base station and avoid service outage. 
In order to improve the accuracy of this approach, sharing information between different entities in the network will be studied in future work.
We will also exploit this method in order to predict blockages and manage resource allocation. 
{Eventually, this work could be further extended to cover mobile UEs and to consider multiple moving blockers}.

\section*{Acknowledgment}
This work was supported by the french government under the Recovery Plan (CRIIOT project).

\bibliographystyle{ieeetr}
\bibliography{biblio.bib}

\end{document}